\documentclass[preprintnumbers,amsmath,amssymb,usenatbib,nofootinbib,superscriptaddress,showkeys,showpacs,11pt]{revtex4-1}

\usepackage{amsmath}
\usepackage{amsfonts,color}
\usepackage{amssymb,float}
\usepackage{graphicx}
\usepackage{appendix}
\usepackage[colorlinks]{hyperref}
\usepackage{xcolor}
\usepackage{orcidlink}
\usepackage{epsf}
\usepackage{bm}
\usepackage{epstopdf}
\usepackage{natbib}
\usepackage{lipsum}
\usepackage{multirow}
\usepackage{nicematrix}

\begin{document}

\title{Correspondence between particle creation and dark components interaction in the context of $f(\mathcal{G})$ gravity}

\author{S. Ganjizadeh}
\affiliation{Department of Physics, Ayatollah Amoli Branch, Islamic Azad University, Amol, Iran}

\author{Alireza Amani\orcidlink{0000-0002-1296-614X}}
\email{Corresponding Author: al.amani@iau.ac.ir}
\affiliation{Department of Physics, Ayatollah Amoli Branch, Islamic Azad University, Amol, Iran}
\affiliation{Innovation and Management Research Center, Ayatollah Amoli Branch, Islamic Azad University, Amol, Mazandaran, Iran}

\author{M. A. Ramzanpour}
\affiliation{Department of Physics, Ayatollah Amoli Branch, Islamic Azad University, Amol, Iran}
\affiliation{Innovation and Management Research Center, Ayatollah Amoli Branch, Islamic Azad University, Amol, Mazandaran, Iran}

\date{\today}

\begin{abstract}
In this paper, we explore the particle creation scenario in the context of $f(\mathcal{G})$ gravity in flat-FLRW metric. For this purpose, from the perspective of thermodynamics and considering an adiabatic universe, we obtain the modified continuity equation in terms of the dynamic number of particles $N$. On the other hand, we obtain Friedmann's equations in $f(\mathcal{G})$ gravity and then write down the continuity equations of the components of matter and dark energy, taking into account the interaction between them. In what follows, by establishing a correspondence between the particle production scenario and $f(\mathcal{G})$ gravity, we obtain the cosmological parameters in terms of $N$. After that, we find cosmological solutions using power-law cosmology and compare them with Hubble parameter data. Finally, we fit the current model with Hubble data and plot the best fit in terms of the redshift parameter.

\end{abstract}



\maketitle
\newpage
\section{Introduction}\label{I}

One of the prominent features of Big Bang cosmology is the expansion of the universe, which is used to describe its accelerated expansion. In fact, the increase in the metric distance between the components of the passage of time, which is attributed to the relative distance between the elements of the universe, is called the accelerated expansion of the universe. As we know, part of the expansion is due to the inertia effect, which plays an important role in the early universe. Another part of this expansion is due to an unknown repulsive force that may originate from a cosmological constant or dark energy, which plays an important role in the present universe. Dark energy is a force that fills the space between objects and has a repulsive property that opposes the force of gravity and leads to the accelerated expansion of the universe. Various models for dark energy have been presented, including cosmological constant, scalar fields, holographic, agegraphic, teleparallel, bouncing universe model, tachyon and etc. \cite{Weinberg-1989, Caldwell-2002, Amani-2011, Sadeghi1-2009, Battye-2016, Li-2012, Amani1-2015, Faraoni-2016, Wei-2009, Amani-2015, Nojiri_2007, Li-2004, Campo-2011, Hu-2015, Amani-2013, Morais-2017, Zhang1-2017, Singh-2016, Sadatian-2024, Assolohou-2024, Malik-2024, Battista-2021, Sahni1-2003, Setare-2008, Brito-2015, Amanifarahani-2012, Amanifarahani1-2012, Bhoyar-2017, Chirde-2018, Singh-2018, Nagpal-2018, Wei-2008, Wei-2007}. In fact, the expansion of the universe was recognized and confirmed by measuring cosmic distances, observing cosmic objects beyond the Milky Way galaxy, cosmic microwave background radiation, and large-scale structures of the universe. But despite the abundant evidence for the increasing speed of the universe, due to the incompatibility of these observations with the standard model of cosmology in the framework of general relativity, they attribute an unexpected negative pressure component to it \cite{Riess_1998, Perlmutter_1999, Bennett_2003, Tegmark_2004}. Therefore, changing the pressure in the system causes the balance of the system to be disturbed, so in this paper, by choosing the  $f(\mathcal{G})$ gravity model in the particle creation platform, we establish the balance of the system \cite{Parker-1968, Hawking_1975}. Here, we consider the universe as a thermodynamic system with an adiabatic process, that is, the system has no heat exchange with the outside environment, ${Q}=0$ \cite{ZelDovich-1972, Prigogine-1989, Parker-2012, Cardenas-2012, Fabris-2014, Rashidi-2018, Biswas-2017, Mandal-2020}.

As mentioned, we tried to describe the accelerated expansion of the universe with the help of the particle creation regime. An important issue in the particle creation model is the particle production rate, $\frac{\dot{N}}{N}$, which depends on the energy flow. Also, we use Gauss-Bonnet gravity, which can be an appropriate candidate to describe the universe's evolution from early time to late time \cite{Nojiri1-2005, Nojiri-2005, Guo-2009, Bamba-2010, Garcia-2011, Bamba-2017, Carloni-2017,  Glavan-2020, Odintsov-2020}. In fact, $f(\mathcal{G})$ gravity is a general function of the Gauss-Bonnet term, $\mathcal{G}$, which arises from string theory prediction. Thus, $f(\mathcal{G})$ gravity is able to adapt to recent observational data for the accelerated universe and to transition from deceleration to acceleration \cite{Chern-1944, Baojiu-2007}.
We note that with the help of $f(\mathcal{G})$ gravity, we calculate the energy density of dark energy, which leads to obtaining its equation of state (EoS). Also, we calculate the energy density of matter with the help of particle creation and using the interaction model between the components of the universe, we obtain the universe EoS. Finally, we use the Hubble data and do the best fit, so that we plot the graphs.
The interesting point of this paper is, to comprehend the concept of  dark matter, we use the particle creation model, which leads to a better understanding of the early universe. And, to realize the concept of dark energy, we use $f(\mathcal{G})$ gravity, which leads to a better cognition of the late universe.

The outline of the current paper is organized as follows:

In Sec. \ref{II}, we review the foundation of $f(\mathcal{G})$ gravity in the flat-FLRW metric. In Sec. \ref{III}, we consider the origin of the matter creation in the universe as a interacting model between the dark parts of the universe. In Sec. \ref{IV}, we obtain the cosmological solutions respecting Hubble parameter data. We consider a specific form of $f(\mathcal{G})$ to solve the model and then plot the cosmological parameters in terms of the redshift parameter. Also, we explore stability analysis. Finally, in Sec. \ref{VI}, we provide a summary of the current model.


\section{Foundation of $f(\mathcal{G})$ gravity}\label{II}

In this section, we start with the following action for the modified Gauss-Bonnet gravity as $f(\mathcal{G})$ gravity in the form
\begin{equation}\label{eqa}
S = \int {d^4x \sqrt{-g} \left(\frac{1}{2\kappa^2} R + f(\mathcal{G})+ L_{m}\right)},
\end{equation}

where $\kappa^2=8\pi G$, and $f$ is an arbitrary and differentiable function of $\mathcal{G} = R^2 - 4 R_{\alpha \beta} R^{\alpha \beta} + R_{\alpha \beta \gamma \delta} R^{\alpha \beta \gamma \delta}$. $R$, $R_{\alpha \beta}$, and $R_{\alpha \beta \gamma \delta}$ are the Ricci scalar, Ricci tensor, and Riemann tensor, respectively, and $L_m$ is a matter Lagrangian density that depends on the space-time metric $g_{\mu \nu}$. In that case, by taking the variation of the action \eqref{eqa} with respect to $g^{\mu \nu}$, we will have the Einstein equation in the following form:
\begin{eqnarray}\label{eq3}
&\frac{1}{2\kappa^2} G^{\mu \nu} - \frac{1}{2}g^{\mu \nu} f(\mathcal{G}) + 2 f_{\mathcal{G}}RR^{\mu \nu} - 4f_{\mathcal{G}}R^{\mu} _{\rho} R^{\nu \rho} + 2f_{\mathcal{G}} R^{\mu \rho \sigma \nu} R_{\rho \sigma} - 2 \left(\nabla^{\mu} \nabla^{\nu} f_{\mathcal{G}} \right) R \notag\\
&+ 2g^{\mu \nu}(\nabla^2 f_{\mathcal{G}}) R + 4 \left(\nabla_{\rho} \nabla^{\nu} f_{\mathcal{G}}\right) R^{\nu \rho} + 4 \left(\nabla_{\rho} \nabla^{\nu} f_{\mathcal{G}} \right) R^{\mu \rho} - 4 \left(\nabla^2 f_{\mathcal{G}}\right)R^{\mu \nu} \notag\\
&- 4 g^{\mu \nu} \left(\nabla_{\rho}  \nabla_{\sigma} f_{\mathcal{G}}\right)R^{\rho \sigma} + 4 \left(\nabla_{\rho}  \nabla_{\sigma} f_{\mathcal{G}}\right)R^{\mu \rho \nu \sigma} = T^ {\mu \nu},
\end{eqnarray}
where $G^{\mu \nu}= R^{\mu \nu}-\frac{1}{2}g^{\mu \nu} R$ is the Einstein tensor, and $T_{\mu \nu}$ is the matter energy-momentum tensor. The background geometric is considered as the flat-FLRW metric in the form
\begin{equation}\label{eqc}
ds^2 = -dt^2 + a^2(t) \left(dx^2 + dy^2 + dz^2\right),
\end{equation}
where $a(t)$ is the scale factor. So, the energy-momentum tensor is written as
\begin{equation}\label{eqd}
T_{\mu \nu} = (\rho_{eff} + p_{eff}) u_\mu u_\nu - (p_{eff})\, g_{\mu \nu},
\end{equation}
where $\rho_{eff}$ is the effective energy density, and $p_{eff}$ is the effective pressure inside the universe. However, the non-zero components of the energy-momentum tensor are as follows (the velocity 4-vectors are introduced as $u^\mu u_\nu = -1$)
\begin{equation}\label{eqe}
T^0_0 = \rho_{eff}, \,\,\,\,\, T^i_i = - p_{eff}, ~~ (i=1,2,3).
\end{equation}

Now, Friedmann's equations are obtained by inserting the above equations into equation \eqref{eqa} as follows: 
\begin{subequations}
\begin{eqnarray}\label{eq5}
&\rho_{eff} = \frac{3}{\kappa^2} H^2 - \mathcal{G} f_{\mathcal{G}} + f(\mathcal{G}) + 24 \dot{\mathcal{G}} H^3 f_{\mathcal{G}\mathcal{G}},\label{eq5-1}\\
& p_{eff} = -8 H^2 \ddot{f_{\mathcal{G}}} - 16 H (\dot{H}+H^2) \dot{f_{\mathcal{G}}} - \frac{1}{\kappa^2} (2\dot{H}+3H^2) - f + \mathcal{G} f_{\mathcal{G}},\label{eq5-2}
\end{eqnarray}
\end{subequations}
where $f_\mathcal{G} = d f/d\mathcal{G}$, $H = \dot{a}/a$ is the Hubble's parameter, and the dot index shows the derivation with respect to cosmic time. Then the Gauss-Bonnet term, $\mathcal{G}$, and Ricci scalar, $R$, are given by
\begin{subequations}
\begin{eqnarray}\label{eq7}
&\mathcal{G} = 24 H^2 (\dot{H}+ H^2),\label{eq7-1}\\
& R=6(\dot{H}+ 2H^2).\label{eq7-2}
\end{eqnarray}
\end{subequations}

Then the effective continuity equation is
\begin{equation}\label{eqh}
{\dot \rho_{eff}} + 3 H \left({\rho_{eff}} + {p}_{eff} \right) = 0.
\end{equation}

In this work, we consider that the universe is dominated by matter and dark energy, represented by
\begin{subequations}
\begin{eqnarray}\label{eqh1}
&\rho_{eff} = \rho_{m} + \rho_{de},\label{eqh1-1}\\
&p_{eff}=p_{m} + p_{de},\label{eqh1-2}
\end{eqnarray}
\end{subequations}
where the indices $m$ and $de$ represent matter and dark energy, respectively. Friedman's equations are then given by
\begin{subequations}
\begin{eqnarray}\label{eqh2}
&\rho_{de} = \frac{3}{\kappa^2} H^2 - \mathcal{G} f_{\mathcal{G}} + f(\mathcal{G}) + 24 \dot{\mathcal{G}} H^3 f_{\mathcal{G}\mathcal{G}} - \rho_{m},\label{eqh2-1}\\
&p_{de} = -8 H^2 \ddot{f_{\mathcal{G}}} - 16 H (\dot{H}+H^2) \dot{f_{\mathcal{G}}} - \frac{1}{\kappa^2} (2\dot{H}+3H^2) - f + \mathcal{G} f_{\mathcal{G}} - p_{m}.\label{eqh2-2}
\end{eqnarray}
\end{subequations}

The continuity equation for the matter and dark energy components with interaction between them is written as
\begin{subequations}\label{eqi}
\begin{eqnarray}
&\dot{\rho}_{m} + 3 H (\rho_{m}+p_m) = \mathcal{Q},\label{eqi1}\\
&\dot{\rho}_{de} + 3 H (\rho_{de}+p_{de}) = -\mathcal{Q},\label{eqi2}
\end{eqnarray}
\end{subequations}
where $\mathcal{Q}$ is the interaction term that represents the energy flow between matter and dark energy. Note that the quantity $\mathcal{Q}$ should be positive, so the second law of thermodynamics is realized \cite{Pavon_2009}, i.e. energy transfer is from dark energy to dark matter. We assume $\mathcal{Q} = 3 b^2 H \rho_m$, where $b^2$ is the coupling parameter or transmission intensity. In that case, we immediately obtain $\rho_m$ in terms of the scale factor from Eq. \eqref{eqi1} as
\begin{equation}\label{eqii}
\rho_m = \rho_{m_0} a^{-3(1-b^2+\omega_m)},
\end{equation}
where $\rho_{m_0}$ represents the current value for matter energy density, as well as $\omega_m=p_m/\rho_m$ is the matter EoS. However, we can obtain the quantity of the matter density parameter in the following form
\begin{equation}\label{Eqiii}
\Omega_{m} = \frac{\rho_{m}}{\rho_c},
\end{equation}
where $\rho_c = 3 H^2 / \kappa^2$ is the critical density, and the present value of the corresponding quantity is
\begin{equation}\label{Eqiv}
\Omega_{m_0} = \frac{\kappa^2 \rho_{m_0}}{3 H_0^2}.
\end{equation}

Therefore, the dark energy EoS is calculated as
\begin{equation}\label{eqj}
\omega_{de} = \frac{p_{de}}{\rho_{de}},
\end{equation}
where it depends on the dark parts of the universe and Gauss-Bonnet gravity.

\section{Matter creation as the interacting model in the universe}\label{III}

In this section, we intend to explore the issue of matter creation in the expanding universe. It should be noted that this process occurred during the inflationary period of the early universe and was so strong that matter was created through particle production mechanisms. Therefore, by inspiring this mechanism, we assume that particle production has also occurred in the current universe, and this causes cosmic expansion to occur. For this purpose, we review the first law of thermodynamics, including particle production, and consider it as an open system. In that case, one reads
\begin{equation}\label{eqk}
dE = dQ - p_m dV + \frac {h}{n} d(n V),
\end{equation}
where $h = \rho_m + p_m$ is the enthalpy density, $N$ is the dynamic number of particles, $\rho_m = E/V$ and $n = N/V$ are the matter energy density and the density of the particle number, respectively, in which $E$ and $V$ are the internal energy and the volume of the system, respectively. Now, in order to relate the thermodynamic system to the current universe, we consider the corresponding system as adiabatic, because there is no heat exchange from the inside of the universe to its outside. In that case, we consider the thermodynamic system of the universe as a sphere whose radius is the scale factor and its volume is $V = 4\pi a^3/3$. So, we have
\begin{equation}\label{eqm}
\dot \rho_m + 3 H (\rho_m + p_m) -  \frac {\dot N}{N} (\rho_m+p_m) = 0,
\end{equation}
where this relation is introduced as the modified continuity equation for particle production. We note that the negative pressure causes the accelerated expansion of the universe, and even the third term of this equation will be added to the cause.
Thus, the matter energy density is obtained from the differential equation \eqref{eqm} as 
\begin{equation}\label{eqn}
\rho_m = C \left(\frac{N}{a^3}\right)^{1 + \omega_m},
\end{equation}
where $C$ is an integral constant. 

In what follows, we consider the universe dominated by the dynamic number of particles, so that the mechanism of particle creation causes the formation of the contents of the universe from the early period to the late period. Therefore, with this approach, the mechanism of particle creation can be attributed to the evolution of the universe. In that case, the cosmic structure and the universe's evolution are related to the interaction between the components of the universe, i.e., the interaction between the universe components causes the uniform distribution evolution of baryons in the formation of the universe and causes the instability in the dark parts from the early times to the dominance of dark energy in the current universe \cite{Zimdahl-2001, Dutta-2018}. Therefore, with correspondence between the relations \eqref{eqi1} and \eqref{eqm}, the matter production rate is obtained as follows:
\begin{equation}\label{eqo}
\frac {\dot N}{N} = \frac{3 b^2}{1 + \omega_m} H,
\end{equation}
which is a positive value when the energy flows from matter to dark energy. Immediately, $N$ and $\rho_m$ are obtained in terms of the scale factor in the form
\begin{subequations}
\begin{eqnarray}\label{eqo1}
&N = N_0 \, a^{\frac{3 b^2}{1 + \omega_m}},\label{eqo11}\\
&\rho_m = C \, a^{-3 \left(1-b^2+\omega_m \right)},\label{eqo12}
\end{eqnarray}
\end{subequations}
where we consider $N_0=1$ as the present number of particles and also $C=\rho_{m_0}$ which comes from Eq. \eqref{eqii}. It should be noted that in the absence of the interaction model ($b = 0$) and matter ($\omega_m = 0$), dust-like matter dominates the universe. Finally, we can write down the matter energy density in terms of $N$ as follows:
\begin{equation}\label{eqo2}
\rho_m = \rho_{m_0} N^{-\frac{\left(1+\omega_m \right) \left(1-b^2+\omega_m \right)}{b^2}}.
\end{equation}

\section{Cosmological solutions based on Hubble parameter data}\label{IV}

In this section, we intend to solve the particle production mechanism in the context of $f(\mathcal{G})$ gravity. However, we consider the universe to be adiabatic because there is no heat exchange with the outside universe, but we consider the production of particles within the universe that results from the interaction between matter and dark energy. This issue is observed in the continuity equation which includes the interaction term $\mathcal{Q}$. Now, to continue this process, we need to reconstruct the Friedmann equations in terms of the redshift parameter $z$. Hence, we express the relationship between $z$ and $a$, and its differential form is as follows:
\begin{subequations}\label{eqp}
 \begin{eqnarray}
 &a(t) = \frac{a_0}{1+z}, \label{eqp-1}\\
 &\frac{d}{dt} = - H (1+z) \frac{d}{dz}, \label{eqp-2}
\end{eqnarray}
\end{subequations}
where $a_0=1$. In what follows, we opt for the specific model, so-called, the power-law cosmology based on its ease of use as
\begin{equation}\label{eqq}
a(t) = {a_0} {\left(\frac{t}{t_0}\right)^n},
\end{equation}
where $t_0$ and $n$ are the present age of the universe and dimensionless positive coefficient, respectively. Immediately obtains in the form
\begin{subequations}\label{eqr}
 \begin{eqnarray}
 &H = \frac{n}{t}, \label{eqr-1} \\
 &t_0 = \frac{n}{H_0}, \label{eqr-2} \\
 &H(z)=H_0  (1+z)^{\frac{1}{n}}, \label{eqr-3}
\end{eqnarray}
\end{subequations}
where $H(z=0)=H_0=67.4 \pm 0.5 \, km\,s^{-1}\,Mpc^{-1}$ is the current value of Hubble parameter \cite{Aghanim-2017}. In order to survey the value of $n$, we have to fit Eq. \eqref{eqr-3} to the Hubble parameter data, the corresponding result of which can be found in Ref. \cite{Moresco_2016, Farooq_2017, Pacif_2017, Magana_2018, pourbagher1-2020, Mahichi-2021} as $n = 0.956$ and $t_0 = 13.75\, Gyr$. Therefore, we can relate $N$ to $z$ by using Eqs. \eqref{eqo11} and \eqref{eqp-1} in the following form
\begin{equation}\label{eqs}
N = (1+z)^{\frac{-3 b^2}{1 + \omega_m}},
\end{equation}
where $H$, $\dot{H}$, $\mathcal{G}$, and $\mathcal{\dot{G}}$ obtain in terms of $N$ as
\begin{subequations}
\begin{eqnarray}\label{eqt}
&H = H_0 N^{\frac{-(1 + \omega_m)}{3 n b^2}},\label{eqt-1}\\
&\dot{H} = -\frac{H_0^2}{n} N^{\frac{-2(1 + \omega_m)}{3 n b^2}},\label{eqt-2}\\
&\mathcal{G} = -\frac{24 \left(1-n \right) H_0^4}{n} N^{-\frac{4 \left(1+\omega_m\right)}{3 b^{2} n}},\label{eqt-3}\\
&\mathcal{\dot{G}} = \frac{96 \left(1-n \right) H_0^5}{n^2} N^{-\frac{5 \left(1+\omega_m\right)}{3 n b^{2}}},\label{eqt-4}\\
&\dot{f}_{\mathcal{G}} = \frac{3 H_0 b^2}{1+\omega_m} N^{\frac{3 n b^2-\omega_m-1}{3 n b^{2}}} f_{\mathcal{G}N},\label{eqt-5}\\
&\ddot{f}_{\mathcal{G}} = \frac{3 H_0^2 b^2}{n (1+\omega_m)^2} \left((3 n b^2-\omega_m-1) N^{\frac{3 n b^2-2 \omega_m-2}{3 n b^{2}}} f_{\mathcal{G}N} + 3 n b^2 N^{\frac{2(3 n b^2-\omega_m-1)}{3 n b^{2}}} f_{\mathcal{G}NN}\right),\label{eqt-6}
\end{eqnarray}
\end{subequations}
where $f_{\mathcal{G}N} = df_{\mathcal{G}}/dN$ and $f_{\mathcal{G}NN} = df_{\mathcal{G}N}/dN$.

Now, in order to proceed to solve the cosmology, we insert Eqs. \eqref{eqo2} and \eqref{eqt} into Eqs. \eqref{eqh2} and we have
\begin{subequations}\label{equ}
\begin{eqnarray}
&\rho_{de} = \frac{3 H_0^2}{\kappa^2} N^{-\frac{2 \left(1+\omega_m\right)}{3 n b^{2}}} + \frac{24 \left(1-n \right) H_0^4}{n} N^{-\frac{4 \left(1+\omega_m\right)}{3 n b^{2}}} f_{\mathcal{G}} + \frac{2304 \left(1-n \right) H_0^8}{n^2} N^{-\frac{8 \left(1+\omega_m\right)}{3 n b^{2}}} f_{\mathcal{G}\mathcal{G}} + f \notag\\
&- \rho_{m_0} N^{-\frac{\left(1+\omega_m \right) \left(1-b^2+\omega_m \right)}{b^2}},\label{equ-1}\\
&p_{de} = -\frac{24 H_0^4 b^2}{n (1+\omega_m)^2} \left(3 n b^2+(2 n-1)(\omega_m+1)\right) N^{\frac{3 n b^2-4 \omega_m-4}{3 n b^{2}}} f_{\mathcal{G}N} -\frac{72 H_0^4 b^4}{(1+\omega_m)^2} N^{\frac{2(3 n b^2-2 \omega_m-2)}{3 n b^{2}}} f_{\mathcal{G}NN}\notag\\
&- \frac{(3 n-2) H_0^2}{\kappa^2 n} N^{-\frac{2(1+\omega_m)}{3 n b^2}} - \frac{24 (1-n) H_0^4}{n} N^{-\frac{4(1+\omega_m)}{3 n b^2}} f_{\mathcal{G}} - f - \omega_m \rho_{m_0} N^{-\frac{\left(1+\omega_m \right) \left(1-b^2+\omega_m \right)}{b^2}},\label{equ-2}
\end{eqnarray}
\end{subequations}
where dark energy EoS is obtained by the proportionality between the pressure and the energy density of dark energy in the form of $\omega_{de} = p_{de}/\rho_{de}$ (which is not written because it is too long).

Now, for solving the above Friedmann equations, we consider the specific form of the function $f(G)$ from Refs. \cite{Goheer-2009, Rastkar-2012, Munyeshyaka-2023} in the following general form 
\begin{equation}\label{eqv}
f(\mathcal{G}) = C_1 \mathcal{G} + C_2 \sqrt{\alpha \mathcal{G}} + C_3 \mathcal{G}^m,
\end{equation}
where $C_1$, $C_2$, $C_3$, $\alpha$ and $m$ are the constant coefficients. We note that the modified Gauss-Bonnet gravity is reduced to the Gauss-Bonnet gravity when $C_2=C_3=0$. Although the third term plays the role of the first and second terms, our opinion is that the function $f(\mathcal{G})$ is considered as a polynomial in which three terms participate simultaneously, which gives the motivation that it can be more compatible with astronomical data. In what follows, the function $f$ and the corresponding derivatives obtain in terms of $N$ as
\begin{subequations}\label{eqw}
\begin{eqnarray}
&f = -\frac{24 C_1 \eta}{n} N^{\frac{-4 \left(1+\omega_m\right)}{3 n b^{2}}} + 2 C_2 \sqrt{\frac{-6 \alpha \eta}{n}} N^{\frac{-2 \left(1+\omega_m\right)}{3 n b^{2}}} + C_3 \left(\frac{-24 \eta}{n}\right)^m N^{\frac{-4 m \left(1+\omega_m\right)}{3 n b^{2}}},\label{eqw-1}\\
&f_{\mathcal{G}} = C_1 + \frac{C_2 \alpha}{2 \sqrt{\alpha \mathcal{G}}} +C_3 m \mathcal{G}^{m-1},\label{eqw-2}\\
&f_{\mathcal{GG}} = -\frac{C_2 \alpha^2}{4 \left(\alpha \mathcal{G}\right)^{\frac{3}{2}}} + C_3 m (1-m) \mathcal{G}^{m-2},\label{eqw-3}\\
&f_{\mathcal{G}N} = \frac{32 \eta (1+\omega_m)}{n^2 b^2} N^{-\frac{4 (1+\omega_m)}{3 n b^2}-1},\label{eqt-4}\\
&f_{\mathcal{G}N N} = -\frac{32 \eta (n b^2+\frac{4}{3}+\frac{4}{3} \omega_m)}{n^3 b^4} N^{-\frac{2(3 n b^2+2+2 \omega_m)}{3 b^2 n}},\label{eqt-5}
\end{eqnarray}
\end{subequations}
where $\eta = H_0^4 (1-n)$. To substitute Eq. \eqref{eqs} into Eqs. \eqref{eqw}, and then into \eqref{equ} we obtain $\rho_{de}$ and $p_{de}$ in terms of the redshift parameter as follows:
\begin{subequations}\label{eqx}
\begin{eqnarray}
&\rho_{de} = \frac{C_3 n^{-m} (m-1) (4 m+1-n) (-24+24 n)^m}{-1+n} H_0^{4m} (1+z)^{\frac{4m}{n}} - \rho_{m_0} (1+z)^{3(1-b^2+\omega_m)} \notag\\
&+ \frac{3 \sqrt{\alpha n (-1+n)} + \sqrt{6} \alpha C_2 \kappa^2 (1+n)}{\kappa^2 \sqrt{\alpha n (-1+n)}} H_0^2 (1+z)^{\frac{2}{n}},\label{eqx-1}\\
&p_{de} = 24^{m} C_3 n^{-m} \left(m -1\right) \left(-1+n \right)^{m} H_0^{4 m} \left(1+z \right)^{\frac{4 m}{n}} - \frac{n \kappa^2 C_2 \sqrt{6 \alpha (-1+n)} - 2 \sqrt{n} + 3 n \sqrt{n}}{\kappa^2 n \sqrt{n}} H_0^2 (1+z)^{\frac{2}{n}}\notag\\
&\frac{768 \left(-1+n \right) \big(-5 + 2 n+\left(2 n -1\right) \omega_m^{2}+ \omega_m \left(3 b^{2} n +4 n -6\right) \big) H_0^{8} \left(1+z \right)^{\frac{8}{n}}}{n^{3} \left(1+\omega_m\right)^{2}} - \omega_m \rho_{m_0} (1+z)^{3(1-b^2+\omega_m)}.\label{eqx-2}
\end{eqnarray}
\end{subequations}

\begin{figure}[h]
\begin{center}
\includegraphics[scale=.35]{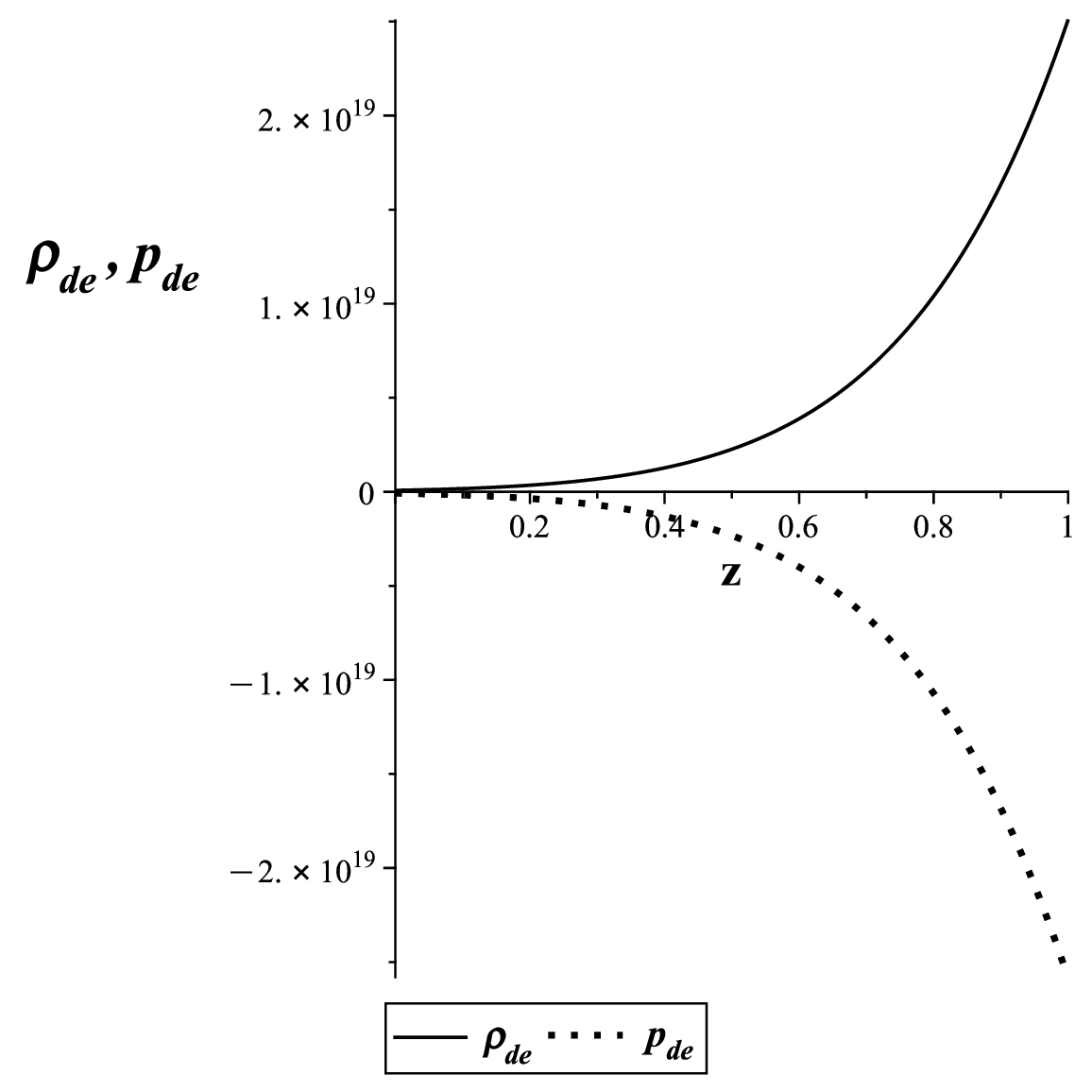}
\caption{The energy density (line) and the pressure (point) of dark energy in terms of redshift parameter for $C_1=C_2=2$, $C_3=-0.8$, $m=2$, $\alpha=-1$, $b=4.5$, $\omega_m=12$, and $\rho_{m_0}=4225$.}\label{Figa}
\end{center}
\end{figure}

Now, to describe the model, we selectively test the current model with some free parameters, which are chosen here as $C_1=C_2=2$, $C_3=-0.8$, $m=2$, $\alpha=-1$, $b=4.5$, $\omega_m=12$, and $\rho_{m_0}=4225$. It should be noted that the corresponding choice is so sensitive and plays an important role in the evolution of the universe. The motivation for this choice is to satisfy the conditions $\rho_{de} > 0$ and $p_{de} < 0$ during the universe periods. For this purpose, we consider a series of answers for the values of free parameters as $C_1=C_2=2$, $C_3=-0.8$, $m=2$, $\alpha=-1$, $b=4.5$, $\omega_m=12$, and $\rho_{m_0}=4225$. In what follows, we plot the variation of energy density and pressure of dark energy as a function of the redshift parameter as shown in Fig. \ref{Figa}. From Fig. \ref{Figa}, we can see that the energy density changes from a very high positive value at an early time (for a large $z$) to a much lower positive value at a late time ($z = 0$), and the pressure starts from a very large negative value at the early time to a much lower negative value at the late time.

\begin{figure}[h]
\begin{center}
\includegraphics[scale=.35]{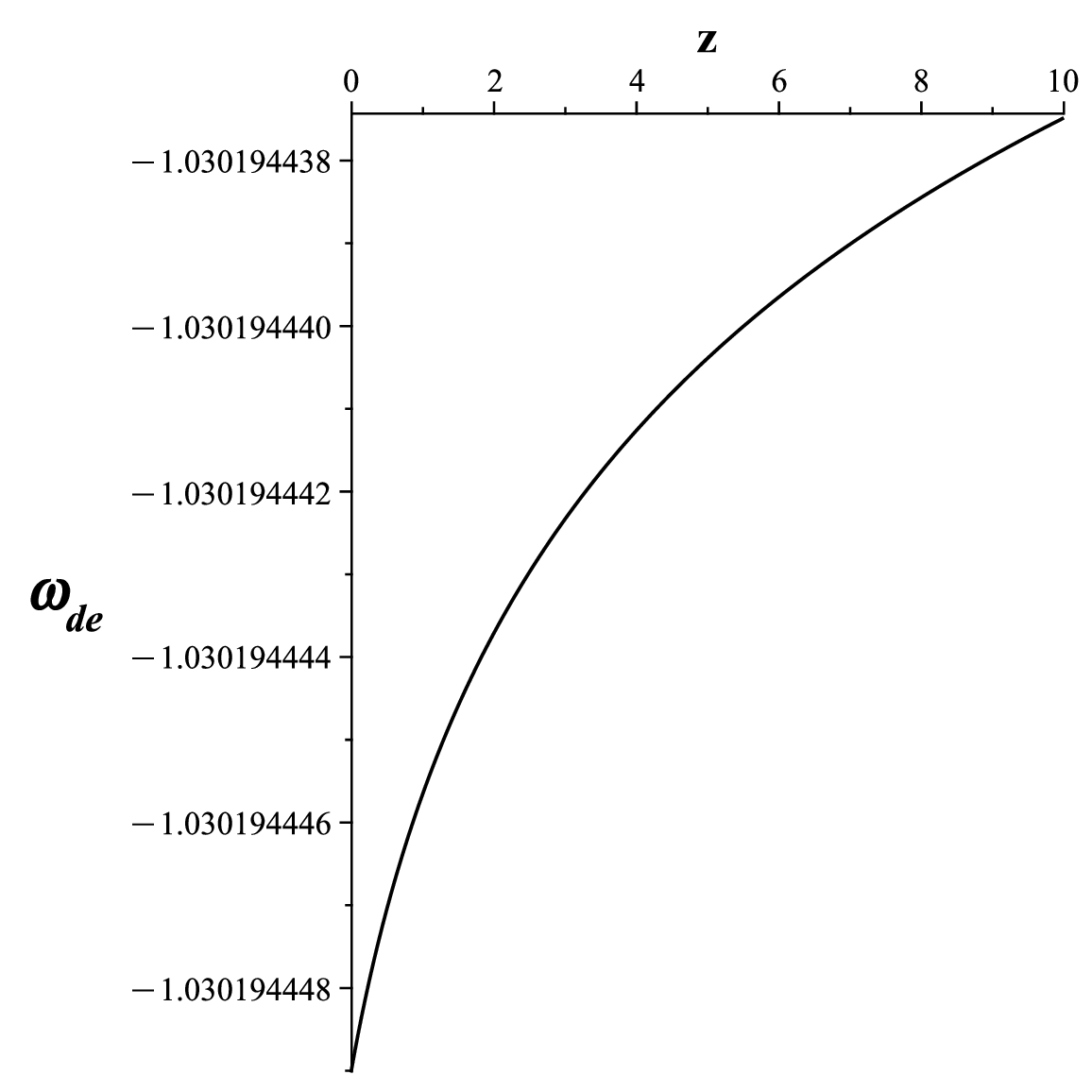}
\caption{The EoS parameter of dark energy in terms of redshift parameter for $C_1=C_2=2$, $C_3=-0.8$, $m=2$, $\alpha=-1$, $b=4.5$, $\omega_m=12$, and $\rho_{m_0}=4225$.}\label{Figb}
\end{center}
\end{figure}

Now to calculate the dark energy EoS parameter, we substitute $\rho_{de}$ and $p_{de}$ from \eqref{eqx} into the Eq. \eqref{eqj}, and then we plot the variation of the redshift parameter as shown in Fig.  \ref{Figb}. The importance of this parameter is that it shows the evolution of the universe from early to late times. We observe from Fig. \ref{Figb} that the value of EoS parameter decreases from early time to late time, in the sense that the evolution of the universe starts from the quintessence epoch ($\omega_{de} > -1$) and then crosses through the phantom epoch ($\omega_{de} < -1$). In fact, the Figs. results confirm that the universe is undergoing an accelerated expansion phase, which agrees with the results of Ref. \cite{Scolnic_2018}.

In what follows, to measure the energy budget within the universe, we must calculate the values of the density parameters for dark energy and matter. For this purpose, we obtain the density parameter of matter in terms of the particle number by inserting Eqs. \eqref{eqo2} and \eqref{eqt-1} into Eq. \eqref{Eqiii} as
\begin{equation}\label{eqy}
\Omega_m = \frac{\kappa^2 \rho_{m_0}}{3 H_0^2} N^{\frac{\left(1+\omega_m \right) \left(2-3n(1-b^2+\omega_m) \right)}{3 b^2 n}},
\end{equation}
where with the help of the Friedmann's first equation, we clearly get $\Omega_{de_0}=1-\Omega_{m_0}$  for the present values. Then, by substituting the values of the above free parameters, we obtain the current values of the density parameters as $\Omega_{m_0}=0.31$ and $\Omega_{{de}_0}=0.69$ as shown in Tab. \ref{taba}. 
Therefore, the energy budget shows us that the results of this study is compatible with the value obtained $\Omega_{m_0} = 0.315 \pm 0.007$ in Ref. \cite{Aghanim-2017}.

\begin{table}[h]
\caption{The present values of the density parameters of $\Omega_{m_0}$ and $\Omega_{de_0}$.} 
\centering 
\begin{tabular}{||c |c ||} 
\hline\hline 
~$\rm{the~ density~ parameter}$~ & $\rm{the~ present~ value}$ \\ [0.5ex] 
 \hline
$\Omega_{m_0}$ &  $0.31$  \\
 [1ex] 
\hline 
$\Omega_{de_0}$ &  $0.69$ \\
 [1ex] 
\hline\hline 
\end{tabular}
\label{taba} 
\end{table}


\section{Conclusion}\label{VI}

In this paper, we examined the evolution of the universe from the early era to the present era, with the help of the particle creation approach in the context of $f(\mathcal{G})$ gravity in the FRW background. The main goal of this study is to be able to study the evolution of the universe depending on the energy flow in it. However, the category of energy flow can be examined from two perspectives. The first is the Friedman equations resulting from modified Gauss-Bonnet gravity, and the second is related to the particle production rate in the universe (which is related to the issue of particle creation).\\
For the first case in the Sec. \ref{II}, we considered action \eqref{eqa}, which includes statements such as the gravitation term, Gauss-Bonnet function $f(\mathcal{G})$ and Lagrangian density of matter. After obtaining Einstein's equation \eqref{eq3} and writing the energy-momentum tensor \eqref{eqd}, we obtained Friedman's equations \eqref{eq5} with background \eqref{eqc}. The effective continuity equation for the entire contents of the universe was calculated and obtained according to Eq. \eqref{eqh}. After considering the contents of the universe in terms of its dark parts, we were able to write their continuity equations separately as Eqs. \eqref{eqi}, taking into account the interacting term $\mathcal{Q} = 3 b^2 H \rho_m$. The interaction expression, $\mathcal{Q}$, is considered based on the energy transfer strength from dark matter to dark energy, which is a conventional quantity derived from phenomenology. One of the important foundations of the paper in this field is related to the interaction term, which we consider proportional to particle production rate, and we will have a more explicit statement about it a little later. Now, in order to solve Friedman's equations, we are only able to solve Eq. \eqref{eqi1}, the answer to which is Eq. \eqref{eqii}. The obtained equation \eqref{eqii} shows the relationship between the atter energy density in terms of scale factor and coupling parameter.\\
Next, the second case has been discussed in Sec. \ref{III}. In this section, we investigated the mechanism of particle production in the evolution of the universe, which occurs during the inflationary period. For this purpose, we consider that the particle creation continues until the late universe. Therefore, with these interpretations, we consider the universe as a closed system that has no heat exchange with the outside universe. For this reason, we first wrote a system that includes particle creation using the first law of thermodynamics as Eq. \eqref{eqk}. Then, we considered this relationship with an adiabatic universe system as a sphere to the scale factor radius, which led to the continuity Eq. \eqref{eqm}. Eq. \eqref{eqm} has been introduced as the modified continuity equation, including the particle creation term, which is the cause of the rapid expansion of the universe. The solution of the differential Eq. \eqref{eqk} leads us to Eq. \eqref{eqn}, which leads to the fact that the matter energy density is related to the dynamic number of the particle. Therefore, from Eq. \eqref{eqn}, we find that the fate of the universe evolution is related to the particle production mechanism. However, the main key of the paper is to correspond the modified continuity equation for the particle production \eqref{eqm} with the continuity equation of matter resulting from $f(\mathcal{G})$ gravity \eqref{eqi1} and we reach the result \eqref{eqo}, and finally the matter energy density in terms of the particle number $N$ and the coupling coefficient of transmission intensity of energy flow between the dark parts of universe $b^2$ was obtained in the form of Eq. \eqref{eqo2}.\\
In what follows, in order to solve the cosmology of the current model, we consider a special model called power-law cosmology according to Eq. \eqref{eqq} in which $n$ and $t_0$ is introduced as the current age of the universe and the correction factor. To evaluate the current model with observational data, we compared Eq. \eqref{eqr-3} with the Hubble parameter data, and from the result of this comparison, the values of $n=0.965$ and $t_0=13.75~Gyr$ were obtained. We note that to avoid rewriting the statistical section, we used the resulting results from Refs. \cite{pourbagher1-2020, Mahichi-2021}. Next, we obtained the value of $N$ in terms of $z$ and subsequently the values of $H$, $\dot{H}$, $\mathcal{G}$, and $\mathcal{\dot{G}}$ in terms of $N$ and finally using these values we were able to obtain the energy density and pressure of dark energy according to Eqs. \eqref{equ}. After that, in order to obtain the energy density and dark energy pressure, we need to introduce the function $f(\mathcal{G})$ according to equation \eqref{eqv}, which is selected as a general form of the solutions found in Refs. \cite{Goheer-2009, Rastkar-2012, Munyeshyaka-2023}. However, the functions $f$, $f_{\mathcal{G}}$, $f_{\mathcal{GG}}$, $f_{\mathcal{G}N}$, and $f_{\mathcal{G}NN}$ were obtained as a function of the particle number according to equations \eqref{eqw}. So finally, we obtained the energy density and pressure of dark energy in terms of the redshift parameter as in Eqs. \eqref{eqx}. In order to be able to see the results of our model visually, we drew the graphs of energy density, pressure and EoS of dark energy in terms of redshift parameter and observed that the late universe is expanding at an accelerated rate. In addition to all this, in order to calculate the energy budget of the universe, the energy contribution of each component of the universe is shown according to Table \ref{taba} as the values of $\Omega_{m}=0.31$ and $\Omega_{de}=0.69$, so that it agrees with the results of Ref. \cite{Aghanim-2017}.\\
As a final word, the present work can open a new perspective for future work. Among these issues, we can refer to the investigation of the particle creation mechanism in the tachyon model. Also, as another work, we can mention the investigation of particle production near the horizon of black holes, which can highlight an interesting interaction between quantum mechanics and cosmology.


\end{document}